\documentstyle[12pt]{article}

\parskip=\medskipamount                 
\def\7#1#2{\mathop{\null#2}\limits^{#1}}        
\def\greaterthansquiggle{\raise.3ex\hbox{$>$\kern-.75em\lower1ex\hbox{$\sim$}}}
\def\lessthansquiggle{\raise.3ex\hbox{$<$\kern-.75em\lower1ex\hbox{$\sim$}}}
\def\Nsquiggle{\raise.3ex\hbox{$N$\kern-.75em\lower1ex\hbox{$\,\sim$}}}

\begin{document}
\def\thefootnote{\fnsymbol{footnote}}
\begin{center}
{~~~}\\[.2in]
{\large\bf {Massive Neutrinos as Probe of Higher Unification
\footnote{Abridged version of invited talks delivered at the
Puri Workshop on"Particle Physics and Cosmology at the Interface"
held in January,1993
and Kazimierz School on "New Physics with New Experiments" held
in May,1993}} }\\[.5in]
{\large\bf { R.N.Mohapatra} }\\[.2in]
{\it Department of Physics,University of Maryland,College Park,Md.20742}\\
{UMD-PP-93-208}\\
{June,1993}\\
\end{center}
\begin{center}
{\bf Abstract}\\
\end{center}

     There are strong indications for neutrino masses and mixings
in the data on solar neutrinos as well as in the observed deficit
of muon neutrinos from the atmosphere. The COBE data and other
analysis of the large scale structure in the Universe also seems to
require a hot component in the universe's
dark matter, which can be interpreted as
a massive neutrino with mass in the few eV range. Implications of
a non-vanishing neutrino mass
 for physics beyond the standard model is discussed. It
is argued that a non-zero neutrino mass is a strong indication
of a new local $B-L$ symmetry of electro-weak interactions .
 In particular, it is noted that
  the simplest picture based on the left-right symmetric
unification with a high scale  for the  $B-L$ symmetry
(as well as the scale of the right-handed gauge boson $W_R$) as
 required by the constraints of grandunification and present LEP data
in a minimal $SO(10)$ model, can
accomodate the solar neutrino data and a weak hot dark matter neutrino
but not the atmospheric neutrino puzzle. A model with low scale
$W_R$ ( in the TeV range) where the Dirac mass of the neutrino arises
at the one loop level can also do the same job.
The low $W_R$ picture can be tested in many rare decay experiments
whereas the minimal SO(10) model can be tested by the $\nu_\mu$ to
$\nu_\tau$ oscillation experiments to be carried out soon. Some non-minimal
scenerios to accomodate all three data are also discussed.

\newpage
\begin{center}
{\bf I.\underline{Introduction}:}
\end{center}

   The standard model of electro-weak interactions is a highly
successful, yet a very unsatisfactory theory and it is an absolute
certainty that there is new physics beyond it which is waiting to
be discovered. One of the tell-tale signs of this new physics is
in the arena of neutrinos where there are quite strong indications
for neutrino masses and mixings from several different experiments.
They are : i) deficit of solar neutrinos now observed in four
different experiments[1]; ii) the depletion of atmospheric muon
neutrinos observed in three different experiments[2]; and iii)
the apparent need for some hot dark matter in the Universe[3].
Let us briefly summarize the experimental results in these areas
and state their implications for neutrino masses and mixings.

{\it 1.1 Solar Neutrino Deficit:}

At present four different experiments involving three different
targets ( $Cl^{37}$, electrons in water, and $Ga^{71}$ )[1]
exposed to solar neutrinos have reported measuring a flux of neutrinos
from the sun which seem to be roughly $1\over{2}$
 to $1\over{3}$ times the expected number now calculated by several
groups using the standard solar model.
Defining $\Phi$ as the flux of neutrinos and denoting by a subscript
the observed and the predicted values for this, the results of the
various experiments are:
$$\Phi_{obs}(Cl)=(2.19\pm 0.25)SNU\;\eqno(1)$$
$$\Phi_{STD}(Cl)=(8.0\pm 3.0)SNU~~~[BP,ref.4]\;\eqno(2)$$
$$\Phi_{STD}(Cl)=(6.4\pm 1.4)SNU~~~[T,ref.5]\;\eqno(3)$$
$${{\Phi_{obs}(KAM)}\over{\Phi_{STD}(KAM)}}=(.48\pm .05\pm .06)~~~[ref.1]
\;\eqno(4)$$
Turning now to the Gallium experiments,the SAGE experiment reports
the result[1]:
$$\Phi_{obs}(Ga^{71})=(58^{+17}_{-24}\pm 14)\;\eqno(5)$$
The Gallex results are[1]:
$$\Phi_{obs}(Ga^{71})=(83\pm 19\pm 8)\;\eqno(6)$$.
The theoretical predictions for Gallium are:
$$\Phi_{STD}(Ga^{71})=(122 to 132 \pm 5 to 7)\;\eqno(7)$$
This discrepancy between theory and experiment is called the solar
neutrino problem. Leaving the temperature of the sun as a free
parameter in the problem also fails to fit all four pieces of data
given below
simultaneously if the neutrino is assumed to be massless[6].
Furthermore, the chlorine data also exhibits a time variation which
is apparently anti-correlated with the number of sunspots. In this
talk, I will ignore this possibly exciting feature of the experiments
and concentrate on the remainder of the puzzle which seems to
require a massive neutrino, assuming of course that there is no
flaw in our understanding of the sun. It is worth pointing out
that detailed study of the problem keeping the core temperature
of the sun as a free parameter[6] has been made and it has been
concluded that all the four results cannot be accomodated without
invoking new properties of the neutrino . Qualitatively, the point
is that the B$^8$ neutrino flux varies like $T_c^{18}$ whereas
the $Be^{7}$ neutrino flux goes like  $T_c^{8}$ and the pp neutrino flux
is almost temperature independent . Therefore by lowering the core
temperature of the Sun, $T_c$ by about $4\%$, one can fit the
Kamiokande data ; but this predicts for Chlorine experiment a
neutrino signal of over $4SNU$'s, which is inconsistent with
data at the $2\sigma$ level.
The exciting implication
of this is that this may be the first indication of new physics beyond
the standard model ; however, a final conclusion must wait for further
data not only from the above on-going experiments but also from
experiments under preparation such as the SNO experiment in Canada,
the Borexino experiment in Gran Sasso , Italy.

   The most straightforward interpretation of the present solar neutrino
data taken literally is that massive neutrinos of different
generations mix among themselves so that a neutrino emitted in a
weak interaction process  not being an eigenstate of the Hamiltonian
evolves into another weak eigenstate (such as $\nu_\mu$ or $\nu_\tau$)
as it travels through space. Since the latter do not interact with
the target material, they remain undetected and the original $\nu_e$
flux gets depleted in the process explaining the puzzle. Assuming
mixing only between two species of neutrinos, it has been determined
that there are two kinds of oscillation solutions: one, where the
$\nu_e$ oscillates on its way to the earth
( called the vacuum oscillation [7])
and another, where the oscillation takes place in the solar core
by a mechanism similar to the refraction of light in a medium [8]
( called the Mikheyev-Smirnov-Wolfenstein matter oscillation ).
 For our purpose,what is
important is the values of the mass difference squares of $\nu_e$
and $\nu_\mu$ ( $\Delta m^2\equiv |m^2_{\nu_e}-m^2_{\nu_\mu}|$)
and mixing angle $sin^2{2\theta}$ that fit the four pieces of data
simultaneously. They are as follows:

i) The small angle non-adiabiatic MSW solution:

$$\Delta m^2_{\nu_e \nu_i}\simeq (.3-1.2)\times 10^{-5} eV^2$$
$$ sin^2{2\theta}\simeq (.4-1.5)\times 10^{-2}\;\eqno(8)$$

ii) Large angle MSW solution:

$$\Delta m^2_{\nu_e\nu_i}\simeq (.3-3)\times 10^{-5}eV^2$$
$$sin^2{2\theta}\simeq .6-.9\;\eqno(9)$$

iii) Vacuum oscillation solution:
$$\Delta m^2_{\nu_e\nu_i}\simeq(.5-1.1)\times 10^{-10}eV^2$$
$$sin^2{2\theta}\simeq(.8-1)\;\eqno(10)$$

{\it 1.2 Atmospheric Neutrino Puzzle:}

   In the upper atmosphere,hadronic collisions produce $\pi^\pm$,
which decay to $\mu^{\pm}$ and muon neutrino( or anti-neutrino)
and the $\mu^{\pm}$ subsequently decay to $e^\pm$ and a electron
neutrino ( or $\bar{\nu}_e$ ) and a $\nu_\mu$ ( or a $\bar{\nu}_\mu$).
Thus each pion in its decay produces two muon neutrinos for every
electron neutrino ( or their anti-particles ). Several underground
experiments have looked for the electrons and muons that would
result from the interactions of these atmospheric neutrinos[2] and
in three of the experiments what has been discovered is that the
number of muon events is far less than the number expected on the
basis of the above discussion. In fact, the observed
  flux ratio of muon to electron
events normalized to the flux ratio based on the standard model
, denoted by $R(\mu/e)$ is reported to have values:
$$R(\mu/e)=.60\pm.07\pm .05,~~~[KAMII]$$
$$~~~~~~~=.54\pm.05\pm .12,~~~[IMB]$$
$$~~~~~~~~~~=.55\pm.27\pm .10,~~~[SoudanII]\;\eqno(11)$$

The theoretically expected value of $R(\mu/e)$ is of course one.

    Again, a straightforward explanation of this can be given if
one assumes that $\nu_\mu$ oscillates to another light neutrino.
The data can be fitted with[9] the values of $\Delta m^2_{\nu_\mu
\nu_j}\simeq .5-.005 eV^2$ and $sin^2{2\theta}\simeq .5$. It is
important to point out that there are two other experiments
one by the NUSEX group and one by the Baksan group, which do
not see this effect; they however have lower statistics at this
time and their present results are consistent with the above
estimates of the mass differences and mixing angles.

{\it 1.3 Hot Dark Matter Neutrinos:}

   Data on the extent of structure in the universe is now
available on a wide range of distance scales. Evidence from
the COBE results on the anisotropy of the cosmic microwave
background radiation, galaxy-galaxy angular correlation
' large scale velocity fields, and correlations of galactic
clusters can all be fit[3] by a model of the universe
containing $70\%$ cold dark matter and $30\%$ hot dark matter.
( But perhaps an admixture in the ratio  $90\%$ to $10\%$
of CDM to HDM may not be inconsistent ). It has also been
recently remarked[10] that a warm dark matter alone may be able to
fit the long and short distance parts of the power spectrum.

 A prime candidate for the hot dark matter  is a neutrino
with mass in the appropriate range of $7 eV~ to~ 2 eV$.
 Such a model not only
provides a consistent explanation of the shape of the density
fluctuation spectrum but also the observed estimates of the
absolute density on small and large scales.

{\it 1.4 Other constraints on Neutrinos:}

    Astrophysical and cosmological observations have established
many constraints on the properties of the various types of
neutrinos [11];in this section, we present only those constraints
that are of immediate relevance for the discussion in this
paper. The first is the non-observation of neutrinoless double
beta decay($\beta\beta_{0\nu}$), which provides an important
limit on an effective Majorana mass,$<m_\nu>\leq{1-2}eV$ [12].
Second, if the possibility of light sterile neutrino is considered,
its possible oscillation to one of the active neutrino species
must be severely limited [13] from considerations of nucleosynthesis[14];
which  implies that at the $95\%$ confidence level,the number of
effective neutrino-like relativistic degrees of freedom consistent
with present data on Helium abundance is $\leq{3.3}$. This constrains
the product $\Delta m^2_{\nu_s\nu_i}{sin^4{2\theta}}\leq
 3.6\times 10^{-6}eV^2$.
\bigskip
\begin{center}
{\bf II.\underline{Neutrino Mass Matrices Suggested by Data}:}
\end{center}
\bigskip

   Let us start with a brief reminder about possible structure
of neutrino masses in unified theories. Because of the fact that
the neutrinos do not have electric charge, their mass matrices enjoy
a much richer texture than the charged fermions. It is most convenient
to describe neutrino masses in the language of two component neutrinos
( denoted by $\nu_i$, which transform under an $SL(2,C)$ transformation
{\bf Z} as
$$\nu_i\rightarrow{Z\nu_i}\;\eqno(12)$$

Recalling that {\bf Z} has determinant one in order to qualify as
a Lorentz transformation, a Lorentz scalar can be written as
$i\nu^T_i\sigma_2\nu_j$ where i,j go over all two component chiral
fermions. This statement  holds of course for all fermions with
the important exception that if a fermion has a nonzero "charge",
then in order for the mass term to respect invariance under the
corresponding group of transformations, the mass term above must
necessarily connect two different spinors. Thus the charged
fermions have half the dimensionality of neutral fermions such as
a neutrino. Denoting the elements of the neutrino mass matrix
as $m_{ij}$, it is obvious that for arbitrary mass matrix,there will
be mixings between all neutrinos and therefore neutrino oscillations.

  Let us now proceed to discuss the kind of neutrino spectrum
and their mass matrices that would be required to fit the above
observations.This discussion follows a recent paper by D.Caldwell
and this author[15]. We have isolated three different scenerios
for this purpose (we denote them by A,B,C respectively). In case A,
one can make do only with the three known neutrinos with a fine-tuned
mass matrix whereas cases B and C postulate the existence of one
or more light sterile neutrino species in addition to the three
known weak-interaction-active neutrinos. Let us discuss them separately.

{\it case IIA:}

       In this case, we will assume that $\nu_e$,$\nu_\mu$ and
$\nu_\tau$ are all nearly degenerate with mass around 2 to 2.5
eV. The mass differences are appropriately arranged so that
$\nu_\mu$-$\nu_\tau$ oscillations explain the atmospheric neutrino
problem and similarly $\nu_e$ - $\nu_\mu$ mass differences as well
as mixings are so arranged that they can explain the solar neutrino
deficit via the  MSW mechanism mechanism using the small angle
non-adiabatic solution. The simplest mass matrix, which can achieve
this is:

\renewcommand\arraystretch{1.1}
$$ M = \left( \begin{array}{ccc}
 m & \delta s^2_2 s_1  & -\delta s_1 s_2 \\
\delta s^2_2 s_1  &  m+\delta s^2_2  &  -\delta s_2 \\
-\delta s_1 s_2  &  -\delta s_2  &  m+\delta
\end{array} \right)$$     \hfill (13)

In eq.(13), $m\simeq 2-2.5 eV$; $\delta\simeq .08 eV$; $s_1\simeq .05$
and $s_2\simeq .35$. It is worth pointing out that a value of the
Majorana mass for $\nu_e$ of this magnitude is barely consistent
with the present upper limits from neutrinoless double beta decay
[9] and will be tested in near future by the high precision
$\beta\beta_{0\nu}$ searches using enriched Germanium now under
progress[12]. A very simple way to avoid the double beta decay
constraint would be to assume the three neutrinos to be Dirac
type so that overall lepton number is conserved.
 Obviously,hot dark matter in thus case is, distributed
between the three active species of neutrinos almost equally.

{\it case IIB.}

   In this case, the hot dark matter is identified with one or more
light sterile neutrinos,$\nu_s$[16],whereas the known neutrinos
are assumed to be extremely light with a mass matrix similar in
form to the one in eq.(13) with the important exception that
the value of m is no more in the eV range . As a result, the
naturalness problem in the theoretical derivation of this matrix
is not as severe as in case IIA. The nucleosynthesis constraints
then require the masses of $\nu_s$ to be higher for them to acquire the
mantle of "dark-matter-dom" [15]. The argument goes as follows:
Any light particle present at the epoch of nucleosynthesis will
contribute to the energy density at that epoch and will therefore
effect the Helium abundance. As mentioned before,present data seems
to allow this additional contribution to density to upper bounded
in such a way that its effective contribution is less than that of
.3 neutrino species. Below ,we will denote this parameter by
$\delta N_\nu$ as is conventionally done.
 Since number density scales like the three
quarter power of the energy density,the  number density of these
particles at the era of structure formation is given by:
$$n_{\nu_s}\simeq n_{\nu_e}\times(\delta N_{\nu})^{3\over 4}\;\eqno(14)$$

Its mass ( i.e. $m_{\nu_s}$ ) will therefore be bigger by a factor
$(\delta N_{\nu})^{-{3\over 4}}$. For $\delta N_\nu = .3$, this
leads to $m_{\nu_s}= 17 eV$ for only one species of sterile neutrino.
In actual models with sterile neutrinos, this constraint is
generally implemented by letting the sterile neutrino decouple
much earlier than the epoch of nucleosynthesis.
In many practical situations , the sterile neutrino decouples above
the quark-hadron phase transition temperature $ (\simeq 200-400 MeV)$.
Its contribution to $\delta N_\nu$ then is roughly $\simeq 0.1$. In
such a situation,the required value of the mass, $m_{\nu_s}\simeq 39 eV$.
At the present state of our understanding of the structure formation,
there is enough model dependence so that such high values of $m_{\nu_s}$
cannot perhaps strictly be ruled out; however,they are large enough
so as to be unlikely candidates for hot dark matter.

   Another constraint on such models arise from the fact that, at the
epoch of nucleosynthesis,oscillations to the sterile component fron the
from the
active components must also be suppressed for the same reason as above.
It will therefore be extremely difficult test this scenerio.

{\it Case IIC:}

     In this scenerio,$\nu_{\mu}$ and $\nu_\tau$ are assumed to
have a mass of about 3-4 eV each and constitute the hot dark
matter of the universe and also resolve the atmospheric neutrino
puzzle via their mutual oscillation. The solar neutrino deficit
is explained via the $\nu_e$ to $\nu_s$ oscillation with both
of them having mass in the range of $10{-3}$ eV. Such a picture
was advocated for the 17 keV neutrino by Caldwell and Langacker[17].
The simplest mass matrix for this case is:
 \renewcommand\arraystretch{1.1}
$$ M=\left( \begin{array}{cccc}
\mu_1  &  \mu_3  & 0  &  0 \\
\mu_3  &  \mu_2  & 0  &  0 \\
0  &  0  &  m  & \delta/2 \\
0  & 0  &  \delta/2 &  m+\delta
\end{array} \right) \;\eqno (15)$$

where columns refer to $\nu_s$,$\nu_e$,$\nu_\mu$ and $\nu_\tau$;
$\mu_{1,2}$ are of order $10^{-2}-10^{-3}$ eV; $\mu_3$ is of order
$10^{-4} -10^{-5}$ eV, $m=3-4 eV$ and $\delta=0.07~~ to~~ 0.0007$ eV
in order to accomodate all observations.

    Before ending this section, we like to caution the reader that
the mass matrices displayed in this section are required only if
one takes all three indications of nonvanishing neutrino mass
described above quite seriously. This is by no means absolutely
compelling. In view of this, below we explore the theoretical
implications of only the possible existence of a non-vanishing
neutrino mass rather than any detailed texture for these masses
as for instance envisioned in the above mass matrices. The goal
of such an analysis is not only to expose these plausible models
to tests via neutrino experiments but also to provide
guidance to experimental explorations.

\bigskip
\begin{center}
{\bf III.\underline{ Massive Neutrinos , Local B-L Symmetry,
 and Spontaneous
Parity Violation}:}
 \end{center}
\bigskip
    Let us start by reminding the reader that in the standard model,
the neutrinos are massless because of the fact that only the lefthanded
chirality state of the neutrino appears in the the fermion spectrum
and the fact that $B-L$ is an exact symmetry of the Lagrangian. In order
to obtain massive neutrinos, one must therefore include the right-handed
neutrino in the spectrum. It however turns out that as soon as tn as this
is done, in the theory there appears a completely triangle anomaly free
generator, the $B-L$. This symmetry is then a gaugeable symmetry and
it would be rather peculiar if nature chooses not to gauge a symmetry
which is gaugeable. If following this line of reasoning,we use $B-L$
as a gauge symmetry, the most natural gauge group turns out to be
the Left-Right symmetric group $SU(2)_L\times SU(2)_R\times U(1)_{B-L}$ [18],
which breaks at some high scale to the standard model group. Let us briefly
remind the reader about some of the features of the model.

   Since the gauge group is $SU(2)_L\times SU(2)_R\times U(1)_{B-L}$,this
model has in addition to the usual left-handed $W^\pm_L$ and $Z$, an
additional set of gauge bosons, the $W^{\pm}_R $ and $Z^{\prime}$. The
quarks and leptons in this model are assigned in a completely left-right
symmetric manner,i.e. if we define $Q\equiv (u,d)$ and $\psi \equiv (\nu,
e)$, then $Q_L(2,1,1/3)$ and $Q_R(1,2,1/3)$ are assigned in a left-right
symmetric manner and similarly , $\psi_L(2,1,-1)$ and $\psi_R(1,2,-1)$
( where the numbers in the parenthesis represent the quantum numbers
under the gauge group of the corresponding fields).

The Higgs sector of the model that leads naturally to small
neutrino masses in this model consistss of the bi-doublet field
$\phi\equiv (2,2,0)$ and the triplet fields $\Delta_L\equiv (3,1,+2)$
and  $\Delta_R\equiv (1,3,+2)$. In order to see how the small neutrino
masses arise in this model, let us write down the Yukawa couplings of
the model:
$$L_Y = h_1 \bar{Q}_L\phi Q_R + h_1^{\prime}\bar{Q}_L \tilde{\phi} Q_R
h_{\ell}\bar{\psi}_L \phi \psi_R + h^{\prime}_{\ell}\bar{\psi}_L
\tilde{\phi} \psi_R + f\psi^T_L C^{-1}\tau_2 \Delta_L\psi_L + L\rightarrow
R  + h.c. \;\eqno(16)$$

  The gauge symmetry breaking is achieved in two stages : in the first
stage, the neutral component of $\Delta_R$ multiplet acquires a vev
$<\Delta^0_R>=v_R$, thereby breaking the gauge symmetry down to
the $SU(2)_L\times U(1)_Y$ group of the standard model; in the
second stage, the neutral components of themultiplet $\phi$
acquire vev breaking the standard model symmetry down to $U(1)_{em}$.
At the first stage of symmetry breaking, $W_R$ and $Z^{\prime}$
acquire masses of order $gv_R$ and in the second stage the familiar
$W_L$ and $Z_L$ acquire masses. As mentioned earlier, in order to
understand the near maximality of parity violation at low energies,
the masses of $W_R$ and $Z^{\prime}$ must be bigger than those of the
$W_L$ and the $Z$ bosons. The existing data have been extensively
analysed by various groups[20] and one finds that for the minimal model,
the most model independent limit is provided by the existing LEP data[21]
and corresponds to $M_{Z^{\prime}}\geq 800 GeV$ and using the relation
between the $W_R$ and $Z^{\prime}$ masses present in this model , we get
$M_{W_R}\geq 475 GeV$.

    Now turning to the fermion sector, at the first stage of symmetry
breaking, the f-terms in the Yukawa coupling give nonvanishing masses
to the three right-handed neutrinos of order $fv_R$ keeping all other
fermions massless. At the second stage , quarks, charged
leptons as well as
thethe neutrinos acquire Dirac masses. The $\nu_L$-$\nu_R$ mass mass
matrix at this stage is a $6\times 6$ mass matrix of the following
see-saw form[22]:
\renewcommand\arraystretch{1.1}
$$   M_{\nu}=\left(\begin{array}{cc}
             0 &  m_D \\
             m^T_D& M_R
      \end{array}\right)    \;\eqno(17)$$

   In eq.(17),$m_D$ and $M_R$ are $3\times 3$ matrices.
As is well known this see-saw form leads to three light eigen-values
generically of order
$$m_{\nu_{i}}\simeq \left( {m^2_D}\over {M_R} \right)\;\eqno(18)$$

The typical values of $m_D$ are expected to be of order of the charged
fermion masses in the theory whereas the $M_R$ corresponds to the scale
of $B-L$ breaking which is a very high scale, thereby explaining the
smallness of the neutrino masses. The specific value of $m_D$ is however
model-dependent and depending on what the value of $m_D$ is, the
spectrum of the light left-handed Majorana neutrinos will be of the
eV:keV  : keV : MeV type or of the micro-milli- eV type. The former type
of spectrum can be tested in the double beta decay as well as the
conventional beta decay end-point experiments whereas the second
spectrum can be tested in the solar neutrino as well as the long
base-line neutrino experiments.
In view of the discussion of the previous section,the micro-milli-eV
spectrum is of great current interest. I would therefore like to
explore in the context of left-right symmetric models ways to get
such a spectrum. As mentioned , in the simplest realizations of
these models, the $m_D\simeq m_f$,where $f=$ leptons or quarks.
If we therefore want $m_{\nu_{\mu}}\simeq 10^{-3}$eV,then the mass
$M_R$ must be of order $10^{10}-10^{12}$GeV. This would suggest
grandunification models of type $SO(10)$ or some higher group
containing it. The $SO(10)$ possibility has recently been analysed
by Babu and this author[23] and I present this model in the subsequent
section.

   Another possibility is however that the neutrino Dirac mass
may be radiative in origin in which case it will roughly be
of the magnitude $m_D\simeq (h^2/ 16\pi^2) m_f$. If we then use
the see-saw formula then for $h\simeq .1$ and $M_R\simeq few~~ TeV$,
one can also get a micro-milli-eV type spectrum relevant in the
discussion of solar neutrino puzzle.Such an example was worked
out in ref.[24] and I briefly outline the salient features of
this model. The main lesson to learn from such a model would
be that a neutrino mass in the milli-electron volt range
need not necessarily mean a superheavy scale for $B-L$ symmetry
breaking.

\bigskip
\begin{center}
{\bf IV.\underline{A Low $W_R$ Scenerio for the Solar Neutrino Puzzle}:}
\end{center}
\bigskip

  The particle content and the gauge group of the model is same
the left-right symmetric models with a few additional features:
i) we demand that the model obey a softly broken Peccei-Quinn
symmetry with quarks having PQ charge +1 whereas the leptons
are required to have this charge -1. As a result of this,the
$\phi$ field has PQ charge +2 and we must set $h_1^{\prime}=0$.
On the other hand in the leptonic sector, we must have
$h_{\ell}=0$. The Higgs potential in this case is such that,
the vev of $\phi$ has the form:
\renewcommand\arraystretch{1.1}
$$   <\phi> = \left(\begin{array}{cc}
               \kappa  &  0 \\
                  0  &   0
          \end{array}  \right)$$   \hfill    (19)

 It is then clear that, at the tree level, eq (19) only allows
the up quarks and the charged leptons of the three generations
to pick up mass. The down type quark masses and the neutrino
Dirac masses remain zero at the tree level as do all the CKM
angles. We therefore have to extend the model somewhat to get the
down quark and the neutrino Dirac masses at the one-loop level.
 It was shown  in ref.[24] that this can be achieved by adding to the
model only three extra color triplet scalar fields denoted by
$\omega_L,\omega_R and \omega $. To the Yukawa Lagrangian described
earlier, one can then add the following PQ invariant piece:

$$L^{\prime}_Y=f^{\prime} Q^T_L C^{-1}\tau_2 Q_L \omega_L + L\rightarrow R
\;\eqno(20)$$

In eq.(20),the $f^{\prime}$ is a three by three matrix in generation
space. Note that $\omega_L and \omega_R$ have opposite PQ charges
of $-2$ and $+2$ respectively.
Similarly, to the Higgs potential,we add the term
$\omega^*_L \omega + L\rightarrow R$ (which breaks the PQ invariance
softly)
in addition to the
usual PQ invariant renormalizable terms. It was shown in ref.[24] that
inclusion of the $\omega$ fields and in particular the above soft
PQ breaking terms generate the down quark masses at the one loop
level and also a small but finite vev $\kappa^{\prime}$ that generates
small radiative neutrino Dirac masses as claimed.

Let us present a brief outline of down quark and neutrino Dirac
masses that result in this model:
 $$M_d = \varepsilon f^{\prime} M_u f^{\prime} + r M_u \;\eqno(21)$$

Since the up quark mass matrix at the tree level is diagonal, eqn (21)
should generate not only the down quark masses but also the CKM angles.
Since $f^{\prime}$ is a symmetric matrix, there are 8 parameters
in eqn.(21) and we fit them to reproduce the six unknowns in the
quark sector. In the neutrino
sector, it is the parameter $r$ that determines the Dirac mass
 and in addition, there will be the Yukawa couplings
associated with the $\Delta$ ( i.e. the coupling matrix $f$ )
that enter via the see-saw mechanism:
The neutrino  mass matrix is then given by
$$ M_{\nu} = f v_L + \left({ r^2}\over {v_R}\right) M_{\ell} f^{-1}
M_{\ell} \;\eqno(22)$$

The value of $\kappa^{\prime}$ is estimated to be in the range of
$10 to 50 $ MeV and the value of $v_L$ is of the order of
${\lambda (\kappa^{\prime})^2/ v_R}$ where $\lambda$ is a typical
scalar self coupling ( assumed to be of order $10^{-2}$ ). This
leads to an estimate of $v_L$ of order $\simeq 10^{-1}eV$ or so.
Therefore with an appropriate choice of the parameters $f$ , an
MSW type spectrum can be obtained.

   Thus , while the grand-unified type structure may be the
simplest way to obtain a micro- milli-eV type neutrino spectrum,
a low  $B-L$ scale alternative cannot be completely ruled out.
As already mentioned, such low scale $B-L$ theories can be tested
via the rare decay experiments to be carried out or under way[25].

\bigskip
\begin{center}
{\bf V.\underline{ Minimal SO(10) GUT and Predictions
 for Neutrino Masses and Mixings}:}
\end{center}
\bigskip

The grandunified theories [20](GUTs)provide an elegant extension
of physics beyond the standard model.The  requirement that
the gauge couplings constants in these theories become
equal at the GUT scale ($M_U$) lends them a predictive power
which makes it possible to test them in experiments such as those
looking for the decay of the proton.The most predictive  such
theory is the minimal SU(5) model of Georgi and Glashow , where
the SU(5) symmetry breaks in one step to the standard model.The
only new mass scale in this model is the $M_U$ which can be
determined by the unification requirement using the low energy
values of any two gauge
couplings from the standard model.One then predicts not only
$M_U$, but also the remaining low energy gauge coupling constant
(say $sin^2\theta_W$).It is well known that for the minimal
SU(5) model,they lead to predictions for the proton life-time
as well as $sin^2\theta_W$ both of which are in contradiction
with experiments.

This however does not invalidate the idea of grandunification
and attention has rightly been focussed on SO(10) [26] GUT
models which can accomodate more than one new mass scale.
Supersymmetric SU(5) [27] models also belong to this class.
In this class of two mass scale theories,the values of low
energy gauge coupling constants can determine both the mass
scales again making these theories experimentally
testable.The determination
of the values of the new mass scales become more precise
as the low energy values of the gauge coupling constants
become better known.It is therefore not surprising that
the recent high precision measurement of $\alpha_{strong}$ and
$sin^2\theta_W$ at LEP[28]  once again revived interest in
grandunified theories[29].

Supersymmetric SU(5) theories have
been studied with the goal of predicting the scale of
supersymmetry breaking. These models however donot have
any room for a nonzero neutrino mass nor natural generation
of adequate baryon asymmetry,whereas, the SO(10) model is
the minimal GUT scheme that provides a frame-work for a proper
understanding  both these problems.In  recent papers ,
 SO(10) models with a two step breaking to the standard
model  have been studied[30] and predictions have been given
for the two new mass scales i.e. $M_U$ and $M_I$ making it
possible to predict the proton life-time as well as the
order of magnitude of the neutrino masses. The predictions depend
on the nature of the intermediate symmetry $SU(2)_L \times SU(2)_R
\times G_c$
where $G_c$ is $SU(4)_C$ (denoted as case (A))or
 $SU(3)_cXU(1)_{B-L}$ ( denoted as case (B) ).
     Let us note that in both the above cases
  we have broken the discrete local $Z_2$
subgroup of $SO(10)$ ( called D-parity  in ref.[31]) at the GUT scale
in order to make the see-saw formula natural[32].
Use of Higgs
multiplets belonging to {45} and {210} representations to break SO(10)
leads to such a scenerio[31] automatically.

The predictions for the GUT and Intermediate scale for the
non-SUSY version of this model
have been studied including two-loop and threshold corrections in ref[30]
and the results are:
$$Model (A):~~~~~M_U = 10^{15.8^{+.8}_{-1.7}\pm .2} GeV \newline
{}~~~~~~~~~~~~~~~~~M_I = 10^{11.5^{+2.8}_{-1.5}\pm .02} GeV \;\eqno(23)$$

$$Model (B):~~~~~M_U = 10^{15.8^{\pm .1 \pm .25} GeV \newline
{}~~~~~~~~~~~~~~~~~M_I = 10^{9^{+.6}_{-.3} \pm.18}} GeV \;\eqno(24)$$

 This leads to a prediction for the proton life-time in non-SUSY
$SO(10)$ models for the two chains to be:
$$\tau_p = 1.6\times 10^{35\pm .7\pm .9 ^{+3.2}_{-6.8}} years~~~ Model (A)$$
$$\tau_p = 10^{35 \pm .7\pm 1.0\pm .8} years~~~~Model(B)\;\eqno(25)$$

These predictions are within the reach of the Super-Kamiokande experiment[33],
which should therefore throw light on the non-SUSY version of the
$SO(10)$ model.

Let us now discuss the test of the $SO(10)$ models from neutrino data.
This would require making
precise predictions of the neutrino masses and mixing angles.  This
necessitates  detailed knowledge
 of the Dirac neutrino mass matrix as well as the Majorana neutrino mass
matrix.  Luckily, it turns out that in $SO(10)$ models,
the charge $-1/3$ quark mass matrix is
related to the charged
lepton matrix and
the neutrino Dirac
mass matrix is related
to the charge $2/3$ quark matrix at the unification scale.
However, prior to the work of ref.23 , no somple way was known
 to relate the heavy Majorana matrix to the
charged fermion observables.  This stood in the way of predicting the
light neutrino spectrum. It was however shown in ref.[23] that
 in a class of minimal $SO(10)$
models, in fact, not only the Dirac neutrino matrix, but the Majorana
matrix also gets related to observables in the charged fermion sector.
This leads to a very predictive neutrino spectrum, which we analyze.
We use a simple Higgs system with
one (complex) {\bf 10}
and one {\bf 126} that have Yukawa couplings to fermions.
The {\bf 10} is needed for
quark and lepton masses, the
{\bf 126} is needed for the see--saw mechanism.  Crucial to the
predictivity of the neutrino spectrum is the observation that
the standard model doublet contained in the {\bf 126}
receives an induced vacuum expectation
value (vev) at tree--level.  In its absence, one would
have the asymptotic mass
relations $m_b=m_\tau,~m_s=m_\mu,~m_d=m_e$.
While the first relation would lead to
a successful prediction of $m_b$ at low energies, the last two
are in disagreement with observations.  The induced vev of the standard
doublet of {\bf 126} corrects these bad relations and at the
same time
also relates the Majorana neutrino mass matrix to
observables in the charged fermion sector, leading to a predictive
neutrino spectrum.

We shall consider non--Susy $SO(10)$ breaking to the standard model via
the $SU(2)_L \times SU(2)_R
\times SU(4)_C \equiv G_{224}$ chain as well as
Susy-$SO(10)$ breaking directly to the standard model.
The breaking of $SO(10)$ via $G_{224}$ is achieved by
a {\bf 210} of Higgs which breaks
the discrete $D$--parity[31].
The second stage of symmetry breaking
goes via the {\bf 126}.  Finally, the electro--weak symmetry breaking
proceeds via the {\bf 10}.
In Susy-$SO(10)$, the
first two symmetry breaking scales coalesce into one.

In the fermion sector, denoting
the three families belonging to
{\bf 16}--dimensional spinor representation of $SO(10)$ by
$\psi_a$, $a=1-3$, the complex {\bf 10}--plet of Higgs by $H$, and
the {\bf 126}--plet of Higgs by $\Delta$,
the Yukawa couplings can be written down as
$$
L_Y = h_{ab}\psi_a\psi_bH + f_{ab}\psi_a\psi_b\overline{\Delta} + H.C.
\;\eqno(26)$$
Note that since the {\bf 10}--plet is complex, one other coupling
$\psi_a\psi_b\overline{H}$ is allowed in general.  In Susy--$SO(10)$, the
requirement of supersymmetry prevents such a term.  In the non--Susy
case, we forbid this term by imposing a $U(1)_{PQ}$
symmetry, which may anyway be needed in order to solve the strong CP
problem.

The {\bf 10} and {\bf 126} of Higgs have the following decomposition
under $G_{224}$:
${\bf 126} \rightarrow (1,1,6)+(1,3,10)+(3,1,\overline{10})+(2,2,15)$,
${\bf 10} \rightarrow (1,1,6)+(2,2,1)$.
Denote the $(1,3,10)$ and $(2,2,15)$ components of
$\Delta({\bf 126})$
by $\Delta_R$ and
$\Sigma$ respectively and the $(2,2,1)$ component of $H({\bf 10})$ by
$\Phi$.
The vev $<\Delta_R^0> \equiv v_R \sim 10^{12}~GeV$
breaks the intermediate symmetry down to
the standard model and generates Majorana
neutrino masses given by $fv_R$.
$\Phi$  contains two standard model doublets
which acquire
vev's denoted by $\kappa_u$ and $\kappa_d$ with
$\kappa_{u,d} \sim 10^{2}~GeV$.
$\kappa_u$ generates charge 2/3 quark as well as Dirac neutrino
masses, while $\kappa_d$ gives rise to $-1/3$ quark and charged lepton
masses.

Within this minimal picture, if $\kappa_u,~\kappa_d$ and $v_R$ are
the only vev's
contributing to fermion masses, in addition to
the $SU(5)$ relations $m_b=m_\tau,~
m_s=m_\mu,~m_d=m_e$, eq. (1) will also lead to the unacceptable relations
$m_u:m_c:m_t = m_d:m_s:m_b$.
Moreover, the
Cabibbo-Kobayashi-Maskawa (CKM) mixing matrix will be identity.
Fortunately, within this minimal scheme, we have found
new contributions
to the fermion mass matrices which are of the right order of
magnitude to correct these bad relations.  To see this, note that
the scalar potential contains, among other terms, a crucial term
$$
V_1 =\lambda \Delta \overline{\Delta} \Delta H +H.C.
\;\eqno(27)$$
Such a term is invariant under the $U(1)_{PQ}$ symmetry.  It will be
present in the Susy $SO(10)$ as well, arising from the {\bf 210}
$F$--term.
This term induces vev's for the standard doublets contained in the $\Sigma$
multiplet of {\bf 126}.  The vev arises through a term
$\overline{\Delta}_R\Delta_R \Sigma \Phi$ contained in $V_1$[34].

We can estimate the magnitudes of the induced vev's
of $\Sigma$ (denoted by $v_u$ and $v_d$ along the up
and down directions) assuming the survival hypothesis to hold:
$$
v_{u,d} \sim \lambda \left({v_R^2}\over {M_{\Sigma}^2}\right)
\kappa_{u,d}~~.
\;\eqno(28)$$
Suppose $M_U \sim 10^{15}~GeV$, $M_I \sim 3 \times 10^{12}~GeV$
and $M_{\Sigma} \sim 10^{14}~
GeV$, consistent with survival hypothesis, then $v_u$ and
$v_d$ are of order 100 MeV, in the right range for correcting the bad
mass relations.  We emphasize that there is no need for a
second fine--tuning to generate such induced vev's.  In the Susy
version with no intermediate scale, the
factor $(v_R^2/M_{\Sigma}^2)$ is not a suppression, so
the induced vev's
can be as large as $\kappa_{u,d}$.

We are now in a position to write down the quark and lepton mass
matrices of the model:
$$
M_u = h \kappa_u +fv_u~~$$
$$M_d = h \kappa_d+f v_d$$
$$M_{\nu}^D = h \kappa_u-3 f v_u$$
$$M_l=h \kappa_d-3fv_d$$
$$M_{\nu}^M = f v_R~.
\;\eqno(29)$$
Here $M_{\nu}^D$ is the Dirac neutrino matrix and $M_{\nu}^M$ is the
Majorana mass matrix. Let us ignore CP-violation , which has been
taken into account in [23].

To see the predictive power of the model as regards the neutrino
spectrum, note that we can choose a basis
where one
of the coupling matrices, say $h$, is real and diagonal.  Then there are
13 parameters in all, not counting the superheavy scale $v_R$: 3
diagonal elements of the matrix $h \kappa_u$, 6 elements of $f v_u$, 2
ratios of vev's $r_1=\kappa_d/\kappa_u$ and
 $r_2=v_d/v_u$, and the twophases $\alpha$ and $\beta$.
These 13 parameters are related to the 13 observables in the
charged fermion sector, viz., 9 fermion masses, 3 quark mixing
angles and one CP violating phase.  The light neutrino mass matrix will
then be completely specified in terms of other physical observables
and the overall scale $v_R$.  That would lead to 8
predictions in the lepton sector:  3 leptonic
mixing angles, 2 neutrino mass ratios and 3 leptonic CP violating
phases.

The relations of eq. (4) hold at the intermediate
scale $M_I$ where quark--lepton
symmetry and left--right symmetry are intact.  There are calculable
renormalization corrections to these relations below $M_I$.  The quark and
charged lepton masses as well as the CKM matrix elements run between $M_I$ and
low energies.  The neutrino masses and mixing angles, however, do not
run below $M_I$, since the right-handed neutrinos have masses of order
$M_I$ and decouple below that scale.  The predictions in the neutrino
sector should then be arrived at by first extrapolating the charged fermion
observables to $M_I$.

We shall present results for the non--Susy $SO(10)$ model with the
$G_{224}$ intermediate symmetry.  We fix
the intermediate scale at $M_I = 10^{12}~GeV$ and use the one--loop
standard model renormalization group equations to
 track the running ofthe gauge couplings between $M_Z$ and $M_I$.
For Susy--$SO(10)$, the results are
similar.

To compute the renormalization factors, we choose as low energy inputs
the gauge couplings at $M_Z$ to be
$\alpha_1 (M_Z) = 0.01688,~\alpha_2 (M_Z) = 0.03322,~\alpha_3 (M_Z)
=0.11$.
For the light quark (running) masses, we choose values listed in Ref.
(9).
The top--quark mass will be allowed to vary between 100 and 200 GeV.
Between 1 GeV and $M_Z$, we use two--loop QCD renormalization group
equations for the running of the quark masses and the $SU(3)_C$
gauge coupling,[35] treating
particle thresholds as step functions.  From $M_Z$ to $M_I$, the
running factors are computed semi--analytically both for the fermion
masses and for the CKM angles
by using the
one--loop renormalization group equations for the Yukawa couplings
and keeping the heavy top--quark contribution.[36].  The running factors,
defined as $\eta_i = m_i(M_I)/m_i(m_i)$
[$\eta_i=m_i(M_I)/m_i(1~GeV)$
for light quarks $(u,d,s)$] are $\eta (u,c,t)=(0.273,0.286, 0.506)$,
$\eta (d,s,b)=$ $(0.279,0.279,0.327)$,
$\eta (e,\mu,\tau)=0.960$ for the case of
$m_t=150~GeV$.
The (common) running factors for
the CKM angles (we follow the parameterization advocated by the Particle
Data Group)
$S_{23}$ and $S_{13}$ is 1.081 for $m_t=150~GeV$.
The Cabibbo angle $S_{12}$ and the KM phase
$\delta_{KM}$ are essentially unaltered.

We can rewrite the mass matrices $M_l, M_\nu^D$ and $M_{\nu}^M$ of eq.
(29) in terms of the quark mass matrices and three ratios of
vev's -- $r_1=\kappa_d/\kappa_u,~r_2=v_d/v_u~,R=v_u/v_R$:
$$
M_l = {{4 r_1 r_2}\over{r_2-r_1}} M_u-{{r_1+3 r_2}\over
{r_2-r_1}} M_d ~~,$$
$$M_\nu^D = {{3 r_1+r_2}\over {r_2-r_1}} M_u-{4 \over {r_2-r_1}}M_d ~~,
$$
$$M_\nu^M = {1 \over R} {{r_1} \over {r_1-r_2}}
M_u -{1 \over R}{1 \over {r_1-r_2}}M_d~~.
\;\eqno(30)$$
It is convenient to go to a basis where $M_u$ is diagonal.  In that
basis, $M_d$ is given by $M_d=V M_d^{diagonal} V^T$, where
$M_d^{diagonal}={\rm diagonal}(m_d,m_s,m_b)$ and $V$ is the CKM matrix.  One
sees that $M_l$ of eq. (5) contains only physical observables from the
quark sector and two parameters $r_1$ and $r_2$.  In the CP--conserving
limit then, the three eigen--values of $M_l$ will lead to one mass prediction
for the charged fermions.  To
see this prediction, $M_l$ needs to be diagonalized.  Note first that by
taking the Trace of $M_l$ of eq. (5), one obtains a
relation for $r_1$ in terms of
$r_2$ and the charged fermion masses.  This is approximately
$r_1 \simeq \left(m_\tau+3 m_b\right)/4 m_t$ (as long as
$r_2$ is larger than $m_b/m_t$).
Since $|m_b| \simeq |m_\tau|$ at the
intermediate scale to within 30\% or so, depending on the relative sign
of $m_b$ and $m_\tau$, $r_1$ will be close to either $m_b/m_t$ or
to $(m_b/2m_t)$.
Note also that if $r_2 \gg r_1$, $M_l$ becomes
independent of $r_2$, while $M_{\nu}^D$ retains some dependence:
$M_l \simeq 4 r_1 M_u -3 M_d,~
M_{\nu}^D \simeq M_u-{4 \over {r_2}}M_d$.
This means that the parameter $r_2$ will only be
loosely constrained from the charged fermion sector.

We do the fitting as follows.
For a fixed value of $r_2$, we determine $r_1$ from the Tr$(M_l)$
using the input
values of the masses and the renormalization factors
discussed above.  $M_l$ is then diagonalized
numerically.  There will be two mass relations among charged fermions.
Since the charged lepton masses are precisely known at low energies,
we invert these relations to predict the $d$--quark and $s$--quark
masses.  The $s$--quark mass is sensitive to the muon
mass, the $d$--mass is related to the electron mass.
This procedure is repeated for other values of
$r_2$.  For each choice, the light neutrino masses and the leptonic CKM
matrix elements are then computed using the see--saw formula.

We find that there are essentially three different solutions.  A
two--fold ambiguity arises
from the unknown relative sign of $m_b$ and $m_\tau$ at $M_I$.
Although solutions exist for both signs, we have found that a relative
minus sign tends to result in somewhat large value of $m_s/m_d$.
Our numerical fit shows that the loosely constrained parameter
$r_2$ cannot be
smaller than 0.1 or so, otherwise the $d$--quark mass comes out
too small.  Now, the light neutrino spectrum is sensitive to
$r_2$ only when $r_2 \sim 4 m_s/m_c\sim \pm 0.4$, since the two terms in
$M_\nu^D$
become comparable (for the second family) then.  Two qualitatively
different solutions are
obtained depending on whether $r_2$ is near $\pm 0.4$ or not.

Numerical results for the three different cases are presented below.
The input values of the CKM mixing angles are chosen for all cases to be
$S_{12}=-0.22,~S_{23}=0.052,~S_{13}=6.24 \times 10^{-3}$.
Since $\delta_{KM}$ has been set to zero for now, we have allowed for
the mixing
angles to have either sign.  Not all signs result in acceptable
quark masses though.  Similarly, the fermion masses can have either sign,
but these are also restricted.
The most stringent constraint comes from the $d$--quark
mass, which has a tendency to come out too small.  Acceptable solutions are
obtained when $\theta_{23},~\theta_{13}$ are
in the first quadrant and $\theta_{12}$ in the fourth quadrant.

Solution 1:
$$
{\rm Input}:  m_u(1~GeV) = 3~MeV,~~m_c(m_c)=1.22~GeV,~~m_t =
150~GeV $$
$$m_b(m_b) = -4.35~GeV,~~r_1=-1/51.2,~~r_2=2.0 $$
$${\rm Output}:  m_d(1~GeV) = 6.5~ MeV,~~m_s(1 GeV)=146~MeV $$
$$\left(m_{\nu_e},~m_{\nu_{\mu}},~m_{\nu_\tau}\right) = R\left(2.0 \times
10^{-2},9.9,-2.3 \times 10^4\right)~GeV $$
\renewcommand\arraystretch{1.1}
$$V_{KM}^{\rm lepton} = \left(\begin{array}{ccc}
0.9488 & 0.3157 & 0.0136 \\
-0.3086 & 0.9349 & -0.1755 \\ -0.0681 & 0.1623 & 0.9844\end{array}
\right)$$  \hfill (31)

Solution 2:
$$
{\rm Input}: m_u(1~GeV) = 3~MeV,~~m_c(m_c)=1.22~GeV,~~m_t=150~GeV $$
$$m_b(m_b) = -4.35~GeV,~~r_1=-1/51,~~r_2=0.2 $$
$${\rm Output}: m_d(1~GeV) = 5.6~MeV,~~ m_s(1~GeV)=156~MeV $$
$$\left(m_{\nu_e},m_{\nu_\mu},m_{\nu_\tau}\right) = R\left(7.5 \times
10^{-3},2.0,-2.8 \times 10^3\right)~GeV $$
\renewcommand\arraystretch{1.1}
$$V_{KM}^{\rm lepton} = \left(\begin{array}{ccc}
0.9961 & 0.0572 & -0.0676 \\
-0.0665 & 0.9873 & -0.1446 \\
 0.0584 & 0.1485 & 0.9872 \end{array} \right)$$.
\hfill (32)

Solution 3:
$$
{\rm Input:} m_u(1~GeV) = 3~MeV,~m_c(m_c) = 1.27~GeV,~m_t=150~GeV
$$
$$m_b(m_b) = -4.35~GeV,~~r_1=-1/51.1,~~r_2=0.4$$
$${\rm Output}: m_d(1~GeV) = 6.1~MeV,~~ m_s(1~GeV)=150~ MeV $$
$$\left(m_{\nu_e},m_{\nu_\mu},m_{\nu_\tau}\right) = R\left(4.7 \times
10^{-2},1.4,-5.0 \times 10^3\right)~GeV $$
\renewcommand\arraystretch{1.1}
$$V_{KM}^{\rm lepton} = \left(\begin{array}{ccc}
0.9966 & 0.0627 & -0.0541 \\
-0.0534 & 0.9858 & 0.1589 \\
 0.0633 & -0.1555 & 0.9858 \end{array}  \right)~~$$
\hfill   (33)

Solution 1 corresponds to choosing $r_1 \sim m_b/m_t$.  All the charged
lepton masses are negative in this case.  Since $r_2$ is large, the Dirac
neutrino matrix is essentially $M_u$, which is diagonal; so is the
Majorana matrix.  All the
leptonic mixing angles arise from the charged lepton sector.  Note that
the predictions for $m_d$ and $m_s$ are within the range quoted
in Ref.37.
The mixing angle sin$\theta_{\nu_e-\nu_\mu}$ relevant for solar
neutrinos is 0.30, close to the Cabibbo angle.  Such a value may already
be excluded by a combination of all solar neutrino data taken at the 90\% CL
(but not at the 95\% CL)[6].  Actually, within the model, there is a
more stringent constraint.  Note that the $\nu_\mu-\nu_\tau$ mixing
angle is large, it is approximately $3|V_{cb}| \simeq 0.16$.  For that
large a mixing, constraints from $\nu_\mu-\nu_\tau$
oscillation experiments imply[38] that $|m_{\nu_\tau}^2-
m_{\nu_{\mu}}^2| \le 4~eV^2$.  Solution 1 also has
$m_{\nu_\tau}/m_{\nu_\mu} \simeq 2.3 \times 10^3$, requiring that
$m_{\nu_{\mu}} \le 0.9 \times 10^{-3}~eV$.  This is a factor of 2 too
small for $\nu_e-\nu_\mu$ MSW oscillation for the
solar puzzle (at the 90\% CL), but perhaps is not excluded completely,
once astrophysical uncertainties are folded in.
If $\nu_\tau$ mass is around $2 \times 10^{-3}~eV$,
$\nu_e-\nu_{\tau}$ oscillation may be relevant, that mixing angle is
$\simeq 3|V_{td}| \simeq 6\%$.  It would
require the parameter $R=v_u/v_R \sim 10^{-16}$ or $v_R \sim 10^{16}~
GeV$ for $v_u \sim 1~GeV$.  Such a scenario fits very well within
Susy--$SO(10)$.

Solution 2 differs from 1 in that $r_2$ is smaller,
$r_2=0.2$.  The ratio $m_s/m_d =27.8$ is slightly above the limit in
Ref. 37.  The $1-2$ mixing in the
neutrino sector is large in this case, so it can cancel the
Cabibbo like mixing arising from the charged lepton sector.  As we vary
$r_2$ from around 0.2 to 0.6, this cancellation becomes stronger, the
$\nu_e-\nu_\mu$ mixing angle becoming zero for a critical value of
$r_2$.  For larger $r_2$, the solution will approach Solution 1.
The
$\nu_\mu-\nu_\tau$ mixing angle is still near $3 |V_{cb}|$, so as
before,  $m_{\nu_{\tau}} \le 2~ eV$.  From the $\nu_\tau/\nu_\mu$ mass
ratio, which is $1.4 \times 10^3$ in this case, we see that
$m_{\nu_\mu} \le 1.5 \times 10^{-3}~eV$.  This is just within the
allowed range[6] (at 95\% CL) for small angle non--adiabatic
$\nu_e-\nu_\mu$ MSW oscillation, with a predicted count rate of about
50 SNU for the Gallium experiment.
Note that there is a lower limit of about 1 eV for
the $\nu_\tau$ mass in this case.  Forthcoming experiments
( CHORUS and NOMAD[39] at CERN and the Fermilab expt.) should then
be able to observe $\nu_\mu-\nu_\tau$ oscillations.  A $\nu_\tau$ mass
in the (1 to 2) eV range can also be cosmologically significant, it can
be at least part of the hot dark matter.
In Susy $SO(10)$, $\nu_e-\nu_\tau$ oscillation (the relevant mixing is
about $3|V_{td}| \simeq 5\%$), could account for the solar neutrino puzzle.

Solution 3 corresponds to choosing $r_1 =0.4$.  $m_s/m_d=24.6$ is within
the allowed range.
However, the mass
ratio $\nu_\tau/\nu_\mu$ is $\sim 3.6 \times 10^3$, and
sin$\theta_{\mu \tau} \simeq 3 |V_{cb}|$
so $\nu_e-\nu_\mu$ oscillation
cannot be responsible for solar MSW.  As in other cases,
$\nu_e-\nu_\tau$ MSW oscillation with a 6\% mixing is a viable possibility.

 We have therefore found that contrary to commonly held belief,
 a class of minimal $SO(10)$ grand unified
models can lead to precise predictions for
 light neutrino masses and mixing angles
in terms of observables in the charged fermion sector.  We have been
able to make these predictions within the framework of the minimal
$SO(10)$ model without invoking any family symmetries.

\bigskip
\begin{center}
{\bf VI.\underline{ Summary and Conclusions}:}
\end{center}
\bigskip

       In this report, I have tried to present a brief overview
of the present experimental and theoretical situation with regard to
the neutrino masses . I have taken a dual approach in which one line
of research tries to take all existing evidences for the neutrino
mass seriously and study its implications for the neutrino spectra
and the structure of their mass matrices and their possible theoretical
origin . The other approach is to analyse existing ideas about
new physics beyond the standard model motivated by aesthetic
considerations ( such as fundamental left-right symmetry of nature )
that also imply a non-vanishing neutrino mass to isolate their
experimentally testable predictions . In the later approach,
we have discussed studies of left-right symmetric models with
low scale for the $W_R$ gauge boson , which can not only explain
the solar neutrino puzzle in an appropriate version but also lead
to observable signals in rare decay processes such as muonium-
anti-muonium transition, anomalous muon decay, rare kaon decays etc.
( for a recent review, see [25]). On the other hand , right-handed
scale could be in the super-heavy range of $10^{11}$ GeV or so
( as would be indicated by the simple see-saw mechanism
with a tree level neutrino Dirac mass and the MSW solution to the
solar neutrino puzzle ) . Such high scale theories would arise naturally
in grand-unified models such as those based on the $SO(10)$ group.
Such models can only be tested by the neutrino oscillation experiments
and proton decay searches. In fact, the present
atmospheric neutrino data cannot be accomodated by the minimal $SO(10)$
model; therefore if this data stands the test of time, a second minimal
grandunified model will be ruled out by experiments.

While we have not addressed any GUT theories beyond $SO(10)$, a perfectly
plausible candidate group is the $E_6$ group by Gursey and collaborators
[40]. The fermion spectrum of this model contains a vector-like $SU(2)$
singlet and color triplet fermion, two vector-like $SU(2)$ doublet leptons
and a chiral sterile neutral lepton. It was advocated in [15], that
this sterile lepton can have a very small mass in simple versions
of the model and
may serve the role of a warm dark matter if the MSW mechanism
operates between the $\nu_e$ and $\nu_{\tau}$ species as is expected
in the case of SUSY $SO(10)$. Thus ,the detailed pattern of neutrino
spectrum is indeed a strong clue to the kind of higher symmetries
that are likely to be operative at shorter distances.

We have not addressed issues arising
from recent reports of anomalous events in  double beta decay
experiments involving $^{76}Ge$ , $^{82}Se$ etc.[41].
These reports of unexpected events near the end point of
electron energy could be the first signal of a spontaneously
broken global $B-L$ symmetry [42] as has been discussed in
several recent papers[43]. All these possibilities
promise to keep the area of neutrino physics an extremely
exciting venue for particle physics research in the nineties and
a very optimistic prognosis for the discovery of new physics beyond
 the standard model.

\bigskip
\begin{center}
{\bf Acknowledgement}
\end{center}

The author would like to thank K. S. Babu and D. Caldwell for
 collaborations and discussions; to J. C. Pati and the Institute
of Physics, Bhubaneswar for kind hospitality at Puri, India and
Z. Ajduk and S. Pokorski for kind hospitality at Kazimierz, Poland
where the workshops were held. This work was supported by a grant
from the National science foundation.

\newcounter{000}
\centerline{\bf References}
\begin{list}{[~\arabic{000}~]}
{\usecounter{000}\labelwidth=1cm\labelsep=.5cm}

\item R. Davis et. al., in {\it Proceedings of the 21st International
Cosmic Ray Conference}, Vol. 12, ed. R.J. Protheroe (University of
Adelide Press, Adelide, 1990) p. 143; \\
K.S. Hirata et. al., {\it Phys. Rev. Lett.} {\bf 65}, 1297
(1990); \\
A.I. Abazov et. al., {\it Phys. Rev. Lett.} {\bf 67}, 3332 (1991); \\
P. Anselman et. al., {\it Phys. Lett.}
{\bf B285}, 376 (1992).
\item KAMII.K. Hirata et al.,{\it Phys. Lett.}{\bf B280}, 146 (1992);\\
IMB: R. Becker-Szendy et al., {\it Phys. Rev.}{\bf D 46}' 3720 (1992);\\
Soudan II: M. Goodman , APS-DPF Meeting, Fermilab , November, 1992;\\
Frejus: Ch. Berger et al.,{\it Phys.Lett} {\bf B245}' 305 (1990);\\
Baksan: M.M.Boliev et al., in {\it Proc. 3rd Int. Workshop on
Neutrino Telescopes} (ed. M. Baldo-ceolin, 1991),p.235.
\item E.L.Wright et al.,{\it Astrophys. J.}{\bf 396}, L13 (1992);\\
R. Schaefer and Q. Shafi, BA-92-28 (1992);\\
J. A. Holtzman and J. Primack, {\it Astrophys. J.}{\bf 405}, 428 (1993).
\item J. H. Bahcall and M. Pinsonneault, {\it Rev. Mod. Phys.}{\bf 64},
885 (1992).
\item S. Turck-Chieze et al., {\it Astrophys. J. }{\bf 335} ,415 (1988).
\item S.A. Bludman, N. Hata, D.C. Kennedy and P. Langacker, Pennsylvania
Preprint UPR-0516T (1992);\\
 X. Shi, D.N. Schramm and J. Bahcall, {\it Phys. Rev. Lett.} {\bf 69},
717 (1992).
\item A. Acker, S. Pakvasa and J. Pantaleone, {\it Phys. Rev.}{\bf D43}
, 1754 (1991);\\
V. Barger, R.J.N. Phillips and K. Whisnant {\it Phys. Rev}{\bf D43},
1110 (1991);\\
L. Krauss and S. L. Glashow, {\it Phys. Lett.}{\bf 190B},199 (1987).
\item L. Wolfenstein, {\it Phys. Rev.} {\bf D 17}, 2369 (1978); \\
S.P. Mikheyev and A. Yu Smirnov, {\it Yad. Fiz.} {\bf 42}, 1441 (1985)
[{\it Sov. J. Nucl. Phys}, {\bf 42},
913 (1985)].
\item J. Learned ,  S. Pakvasa and T. Weiler {\it Phys. Lett.} {\bf B207},
79 (1988);\\
V. Barger and K. Whisnant, {\it Phys. Lett.} {\bf 209B},365 (1988);\\
K. Hidaka, M. Honda and S. Midorikawa, {\it Phys. Rev. Lett} {\bf 61},
1537 (1988);\\
W. Frati et al., BA 92-71 (1992).
\item S. Dodelson and L. Widrow, Fermilab Preprint, (1993).
\item R. N. Mohapatra and Palash B. Pal, {\it Massive Neutrinos in
Physics and Astrophysics }, World Scientific, (1991).
\item D. Caldwell et al., {\it Nucl. Phys. ( Proc. Suppl.)} {\bf B13}
547 (1990);\\
A. Balysh et al., {\it Phys. Lett.}{\bf B283}, 32 (1992);\\
For results and the status of more recent experiments, see H.
Klapdor-Kleingrothaus, {\it Nucl. Phys.B ( Proc. Supp.)} {\bf 31}, 72
(1993);\\
R. Brodzinski et al.,{\it ibid.} {\bf 31},76 (1993).
\item R. Barbieri and A. Dolgov, {\it Phys. Lett.} {\bf 237B}, 440 (1990);\\
K. Enquist, K. Kainulainen and M. Thompson, {\it Phys. Lett.}
{\bf B280},245 (1992);\\
J. Cline, OSU Preprint (1991).
\item T. P. Walker et al., {\it Astrophys. J. }{\bf 51},376 (1992).
\item D. Caldwell and R. N. Mohapatra, {\it Phys. Rev.} {\bf D48},xxx (1993).
\item J. Peltoniemi, Valencia Preprint (1993);\\
S. Dodelson and L. Widrow, Fermilab Preprint (1993).
\item D. Caldwell and P. Langacker, {\it Phys. Rev.} {\bf D44},823 (1991).
\item J.C. Pati and A. Salam,{\it Phys.Rev.D$\,$}{\bf 10},275 (1974)\\
R. N. Mohapatra and J.C. Pati,{\it Phys.Rev.D$\,$}{\bf 11},566,2558 (1975)\\
G. Senjanovic and R.N. Mohapatra,{\it Phys.Rev.D$\,$}{\bf 12},1502 (1975).
\item R. N. Mohapatra and G. Senjanovic , {\it Phys. Rev. Lett.}
{\bf 44} , 912 (1980) ; {\it Phys. Rev. }{\bf D23} , 165 (1981).
\item For a review and references, see R. N. Mohapatra,
{\it Unification and Supersymmetry ,second edition}, Springer-verlag (1992);\\
P. Langacker and S. Umashankar, {\it Phys. Rev.} {\bf D40}, 1569 (1989).
\item G. Altarelli et al. {\it Phys. Lett.} {\bf 263B},459 (1991);\\
G. Bhattacharya et al. {\it Phys. Rev.} {\bf D47},3693,(1993);\\
J. Polak and M. Zralek, {\it Phys. Rev.} {\bf D46},xxx, (1992).
\item M. Gell-Mann, P. Ramond and R. Slansky, in {\it Supergravity}, ed.
F. van Nieuwenhuizen and D. Freedman (North Holland, 1979), p. 315; \\
T. Yanagida, in {\it Proceedings of the Workshop on Unified Theory and
Baryon Number in the Universe}, ed. A. Sawada and H. Sugawara, (KEK,
Tsukuba, Japan, 1979);\\
R.N. Mohapatra and G. Senjanovic {\it Phys. Rev. Lett.} {\bf 44}, 912
(1980).
\item K. S. Babu and R. N. Mohapatra, {\it Phys. Rev. Lett.} {\bf 70},
2845 (1993).
\item K. S. Babu and R. N. Mohapatra, {\it Phys. Lett.} {\bf 267B},
400 (1991).
\item for  recent reviews ,see R. N. Mohapatra, {\it Prog.in Part. and
Nucl. Phys.}(to appear) (1994);\\
A. Van der Schaaf, {\it ibid} (1994).
\item H.Georgi,{\it Particles and Fields} ed.C.E.Carlson,A.I.P.,(1975)\newline
H.Fritzsch and P.Minkowski,{\it Ann.Phy.}{\bf 93},193 (1975).
\item S. Dimopoulos and H. Georgi, {\it Nucl. Phys.} {\bf 193 },150 (1981);\\
N. Sakai, {\it Z. Phys. } {\bf C11},153 (1981);\\
R. Arnowitt and P. Nath, {\it Phys. Rev. Lett.} {\bf 69}, 725 (1992);\\
K. Inoue et al. {\it Phys. Rev.} {\bf D45},328 (1992);\\
G. G. Ross and R. Roberts, {\it Nucl. Phys. }{\bf B377}, 571 (1992);\\
J. L. Lopez, D. V. Nanopoulos and H. Pois, {\it Phys. Rev. Lett.}{\bf
70}, 2468 (1993).
\item For a recent summary of LEP data,see T.Hebbekar,Review talk
 at the LEP-HEP conference,Aachen preprint PITHA 91/17.
\item  P.Langacker and M.Luo,{\it Phys.Rev.D$\,$}{\bf 44},817 (1991)\\
 U.Amaldi,W.de Boer and H.Furstenau,{\it Phys.Lett.B$\,$}{\bf 260},447(1991);\\
J.Ellis,S.Kelly and D.V.Nanopoulos,{\it Phys.Lett.B$\,$}{\bf 260},131 (1991).

\item D.Chang,R.N.Mohapatra,J.Gipson,R.E.Marshak and M.K.Parida,
 {\it Phys. Rev. D$\,$}{\bf 31}, 1718 (1985);\\
R. N. Mohapatra and M. K. Parida, {\it Phys. Rev. } {\bf D47}, 264 (1993);\\
N. G. Deshpande, R. Keith and P. B. Pal,{\it Phys. Rev.}{\bf D46}
,2261 (1992).
\item D. Chang, R.N. Mohapatra and M.K. Parida, {\it Phys. Rev. Lett.}
{\bf 52}, 1072 (1982).
\item D. Chang and R.N. Mohapatra, {\it Phys. Rev.} {\bf D 32}, 1248
(1985).
\item Y. Totsuka, KEK Preprint (1992).
\item Possible radiative origin of such mixing terms were noted in
H. Asatryan, Z. Berezhiani and A. Ioannisyan, Yerevan Preprint (1988)
(unpublished);\\
S. Nandi, Private communication.
\item J. Gasser and H. Leutwyler, {\it Phys. Rept.} {\bf 87}, 77 (1982).
\item K.S. Babu, {\it Z. Phys.} {\bf C 35}, 69 (1987);\\
K. Sasaki, {\it Z. Phys.} {\bf C 32}, 149 (1986);\\
B. Grzadkowski, M. Lindner and S. Theisen, {\rm Phys. Lett.} {\bf B198},
64 (1987).
\item D. Kaplan and A. Manohar, Phys. Rev. Lett. {\bf 56}, 2004 (1986).
\item N. Ushida et. al., {\it Phys. Rev. Lett.} {\bf 57}, 2897 (1986).
\item C. Rubbia, CERN Preprint PPE-93-08 (1993).
\item F. Gursey , P. Ramond and P. Sikivie, {\it Phys. Lett.} {\bf 60B},
177 (1976);\\
Y. Achiman and B. Stech, {\it Phys. Lett.} {\bf 77B}, 389 (1977).
\item M. Moe et al. UCI-NEUTRINO-92-1;\\
F. Avignone et al, cited by S. Pakvasa, UH-511-761-93.
\item Y. Chikashige, R. N. Mohapatra and R. D. Peccei, {\it Phys. Lett.}
 {\bf 98B}, 265 (1981).
\item C. Burgess and J. Cline, McGill Preprint (1992);\\
Z. Berezhiani, A. Smirnov and J. W. F. Valle, Valencia Preprint (1992).
\end{list}
\newpage

\end{document}